\documentclass[preprint,12pt]{elsarticle}

\usepackage{graphicx}

\usepackage{amssymb}

\journal{Operations Research Letters}

\begin{document}

\begin{frontmatter}



\title{Approximation of the Quadratic Knapsack Problem}


\author[label1]{Richard Taylor}
\address[label1]{Defence Science and Technology Group, Canberra, ACT 2600, Australia}

\begin{abstract}
For any given $\epsilon>0$ we provide an algorithm for the Quadratic Knapsack Problem that has an approximation ratio within $O(n^{2/5+\epsilon})$ and a run time within $O(n^{9/\epsilon})$.
\end{abstract}


\end{frontmatter}


\section{Introduction}
\label{intro}
The quadratic knapsack problem (QKP) can be defined in graph theoretic terms as follows. Let $G$ be a graph with vertices $v_i, i=1,..,n$ with costs $c_i \geq 0$ associated with the vertices and profits $p_{ij} \geq 0$ associated with each edge $ij$ and each vertex $i$ (corresponding to $p_{ii})$. Given a cost limit $c$ the problem is to find a collection of vertices with total cost at most $c$ with the total profit (the sum of all profits of vertices and edges in the induced subgraph) as large as possible. A particular form of this problem where the costs and profits are all $1$ is equivalent to the densest k-subgraph problem (DkS). Thus DkS asks for the selection of $k$ vertices so that the induced subgraph on those vertices has a maximum number of edges. The reader is referred to \cite{PIS2007} for a survey of the QKP problem, and the recent article \cite{PFE2016} on approximating QKP for special graph classes. In particular we note that \cite{PFE2016} has no results on approximating QKP in the general case, and the author is not aware of any such previously published results. Note that the related problem where negative profits are allowed is known to be inapproximable \cite{RAD2002}.
\section{The Theorem}
\textbf{Theorem.} Let $M$ be any approximation algorithm for DkS that has an approximation ratio within $O(n^{\alpha+\epsilon}), 0<\alpha<1$, $\epsilon>0$ and runs within time $O(n^{1/\epsilon})$. Then $M$ can be used to construct an approximation algorithm $M'$ for QKP that has an approximation ratio within $O(n^{\frac{2\alpha}{1+\alpha}+3\epsilon})$ and runs within time $O(n^{3/\epsilon})$.

Since \cite{BHA2010} provides, for any $\epsilon>0$, an approximation algorithm for DkS with ratio $O(n^{\frac{1}{4}+\epsilon})$ and runs within time $O(n^{1/\epsilon})$ the following corollary follows from the theorem immediately (given $\epsilon$ select $M$ to have approximation ratio $O(n^{\frac{1}{4}+\epsilon/3})$ and run time $O(n^{3/\epsilon})$). \\
\textbf{Corollary.} There is an algorithm for the Quadratic Knapsack Problem that has, for any $\epsilon > 0$, an approximation ratio within $O(n^{2/5+\epsilon})$ and a run time within $O(n^{9/\epsilon})$. \\
\textbf{Proof of the theorem.} Let $G$ be a graph with cost limit $c$, and $M$ an approximation algorithm for DkS that has an approximation ratio of $O(n^{\alpha+\epsilon}), 0<\alpha<1, \epsilon>0$ and runs in time $O(n^{1/\epsilon})$.
\subsection{Overview}
We sketch the main flow of the proof method. We group the costs and profits of the original instance into $O(logn)$ many buckets. This allows us to split the original instance into $O((logn)^3 )$ sub-instances each with a simplified structure. One of these sub-instances has costs and profits on the vertices only, while each of the others have closely bounded costs, no vertex costs and  the same edge profits. In this way we group these sub-instances into five classes: the first is an instance of the classical knapsack problem; while for the other four classes the QKP can be approximated provided we have an approximation for the DkS problem. Finally by a simple averaging argument at least one of the sub-instances must have a profit within a factor of $O((logn)^3 )$ of the original instance. See also \cite{FEI2001} where this argument is used to approximate the weighted DkS problem from approximations of the DkS problem. This paper extends the method further to the QKP.
\subsection{Pruning, Rounding and Grouping} 
We may prune all vertices with $c_i > c$ and all edges $ij$ where $c_i+c_j>c$ since these cannot be part of a feasible solution. Further if $c_i=0$ then the vertex can be pruned and the attached edge values added to the neighbouring vertices without affecting the problem. Also if $p_{ij}=0$ then the edge $ij$ can be pruned. \\
Notation: Let $MAX[G,c]$ be the maximum total profit of the solution to QKP applied to the Graph $G$ with cost limit $c$. Let $G^{MAX}[c]$ be a subgraph of $G$ corresponding to a maximum solution (so the vertex costs of $G^{MAX}[c]$ are at most $c$, and the profits $MAX[G,c]$). Let any subgraph in which the total cost of the vertices is at most $c$ be termed \textit{feasible}. Let $|E(G)|$ denote the number of edges in any graph $G$. 

Now let $p^*$ be the largest $p_{ij}$ and $2^l$ the largest power of $2$ at most $p^*$.  Form the graph $G'$ from $G$ with each $p_{ij}$ rounded down to the nearest profit among $\{2^l,2^{l-1},..,2^{l-q},0\}$ where $q$ is the smallest integer above $log_2n^2=2log_2n$.  Then $MAX[G',c] \geq 1/4MAX[G,c]$ since the effect of deleting all edges with profits less than $2^{l-q}$ and rounding down the remaining profits each reduce the maximum total profit by a factor of at most $1/2$. 
Similarly let $c^*$ be the largest $c_{i}$ and $2^k$ the smallest power of $2$ at least $c^*$. Now group the vertices into buckets $V_i, i=1,..,l+1$ where $l$ is the smallest integer above $log_2n$ and $V_i, i \leq l$ is the collection of vertices $v_j$ where $2^{k-i} < c_j \leq 2^{k+1-i}$ and $V_{l+1}$ is the collection of vertices $v_j$ where $0< c_j \leq 2^{k-l}$.

\subsection{Five subgraph classes} 
The edge and vertex profits of $G'$ can be shared among at most $2(log_2n+1)^3+1$ subgraphs within five classes, with Classes 2-5 constructed based on the cost groupings and profit roundings. We shall show how to obtain $O(n^{\frac{2\alpha}{1+\alpha}+2\epsilon})$ or better approximations for each of these classes in time $O(n^{2/\epsilon})$ or less.\\
Class 1: The subgraph of $G'$ obtained by removing all of the edges. \\
Class 2: Subgraphs with costs within $V_{l+1}$ and edge profits some power of 2, and no vertex profits.\\
Class 3: Subgraphs with costs within $V_{i}, i \leq l$ and edge profits some power of 2, and no vertex profits.\\
Class 4: Bipartite subgraphs with one part having costs within $V_{l+1}$, the other with costs within $V_{i}, i \leq l$, and edge profits some power of 2, and no vertex profits.\\
Class 5: Bipartite subgraphs with one part having costs within $V_{i}, i \leq l$, the other with costs within $V_{j}, i < j \leq l$, and edge profits some power of 2, and no vertex profits.

Class 1 corresponds to the Knapsack problem and so has an algorithm that has an approximation ratio within $1+\epsilon$ and a run time within $O(n^3/\epsilon)$ \cite{VAZ2003}, \cite{GAR1979}. Since $c \geq c^*$ then $2^{k-l} \leq c/n$ so that Class 2 subgraphs have the sum of all vertex costs at most $c$. Thus the maximum edge profit is found by summing all the edge profits. \\
Class 3:  Let $H$ be any subgraph in this class. Let the equal edge profits be $2^x$. We can simplify by scaling the costs and profits so that the vertex costs are between 1 and 2, and the edge profits are all 1. Specifically form a modified subgraph $\tilde{H}$ by dividing the (vertex) costs by $2^{k+i-i}$, and setting the edge profits to $1$. Set the cost limit $\tilde{c}$ for $\tilde{H}$ to $\tilde{c}=c/2^x$. Then
\begin{equation}
MAX[H,c]=\frac{1}{2^{x}}MAX[\tilde{H},\tilde{c}].
\end{equation}
Thus approximating the QKP for $\tilde{H},\tilde{c}$ is equivalent to approximating the QKP for $H,c$. 

We now analyse the graph $\tilde{H}$ with cost limit $\tilde{c}$. Observe that the number of vertices $r$ in any maximum solution must be between $\tilde{c}/2$ and $\tilde{c}$. So let $H^{MAX}[\tilde{c}]$ have $r$ vertices. We may assume $r \geq 4$ or else the problem is trivial. Let $\tilde{H}^t, t=\lfloor \tilde{c}/2 \rfloor$ be any subgraph of $\tilde{H}$ with a maximum number of edges formed by choosing $t$ vertices. In particular we can choose $t$ vertices from $\tilde{H}^{MAX}[\tilde{c}]$ in ${r \choose t}$ ways and the collection of these choices counts each edge of $\tilde{H}^{MAX}[\tilde{c}]$  ${r-2 \choose t-2}$ times. It follows that some such subgraph with $t$ vertices must have a number of edges at least
\begin{equation}
\frac{{r-2 \choose t-2}}{{r \choose t}}\tilde{H}^{MAX}[\tilde{c}]=\frac{t(t-1)}{r(r-1)}\tilde{H}^{MAX}[\tilde{c}] \geq \frac{1}{10}\tilde{H}^{MAX}[\tilde{c}]
\end{equation}
the minimum corresponding to $r=5, t=2$. Thus $\tilde{H}^t$ must have at least as many edges as this subgraph and is also feasible. $\tilde{H}^t$ can be approximated by algorithm $M$ applied to the DkS problem for $\tilde{H},t$ and this provides an $O(n^{\alpha+\epsilon})$ approximation algorithm that runs in time $O(n^{1/\epsilon})$. We note that $\alpha < 2 \alpha/(1+\alpha)$. This completes this class.

In the following for a bipartite graph $H[A,B]$ with parts $A$ and $B$, if $A' \subset A$, $B' \subset B$ then $H[A',B']$ refers to the subgraph of $H[A,B]$ induced by $A'$ and $B'$.\\
Class 4:  Let $H[A,B]$ be any subgraph in this class with parts $A$ and $B$, with the vertex costs of $A$ between 0 and $2^{k-l}$, and the vertex costs of $B$ between $d$ and $2d$, $d$ a power of 2. Since the edge profits are all equal we can scale them as in Class 3, so that they are all 1 in the modified graph $\tilde{H}[\tilde{A},\tilde{B}]$ with cost limit $\tilde{c}$. We may assume $\tilde{c} \geq 4d$ otherwise the problem is essentially finite by considering all choices of at most 4 of the vertices of $\tilde{B}$. For each such choice the choice of corresponding vertices from $\tilde{A}$ amounts to an instance of the Knapsack problem. Similarly we assume $|\tilde{A}| \geq 4$.

Select $\tilde{B}'$ to be the $\tilde{c}/(4d)$ highest degree vertices of $\tilde{B}$ in $\tilde{H}[\tilde{A},\tilde{B}]$. Then $|E(\tilde{H}[\tilde{A},\tilde{B}'])|$ is at least $(1/4)MAX[\tilde{H},\tilde{c}]$ since any feasible solution has at most $\tilde{c}/d$ vertices from $\tilde{B}$. Now select $\tilde{A}'$ to be the $|\tilde{A}|/4$ highest degree vertices of $\tilde{A}$ in $\tilde{H}[\tilde{A},\tilde{B}']$. Then $|E(\tilde{H}[\tilde{A}',\tilde{B}'])| \geq (1/4)|E(\tilde{H}[\tilde{A},\tilde{B}'])|$. Thus $|E(\tilde{H}[\tilde{A}',\tilde{B}'])| \geq (1/16)MAX[\tilde{H},\tilde{c}]$, and since $2^{k-l} \leq \tilde{c}/n$ the subgraph $|E(\tilde{H}[\tilde{A}',\tilde{B}'])|$ is also feasible as the total cost of vertices in each part is at most $\tilde{c}/2$. This completes this class.\\
Class 5:  Let $H[A,B]$ be any subgraph in this class with parts $A$ and $B$. As in Classes 3 and 4 we modify $H[A,B]$ to $\tilde{H}[\tilde{A},\tilde{B}]$ and the cost limit $c$ to $\tilde{c}$, where in $\tilde{H}[\tilde{A},\tilde{B}]$ the edge profits are 1, the vertices in $\tilde{A}$ have costs between 1 and 2, and the vertices of $\tilde{B}$ have costs between $d$ and $2d$, where $d$ is a positive power of 2 as in figure 1.  We may also assume $\tilde{c} \geq 4d$ (see the argument in Class 4). We consider two cases. \\
Case 1: $|\tilde{A}| \leq \tilde{c}^{(1+\alpha)/(1-\alpha)}$: Select the $\tilde{c}/(4d)$ highest degree vertices $\tilde{B}'$ of $\tilde{B}$. Then $|E(\tilde{H}[\tilde{A},\tilde{B}'])| \geq 1/4MAX[\tilde{H},\tilde{c}]$ since any feasible solution can have at most $\tilde{c}/d$ vertices of $\tilde{B}$. Now select the $\tilde{c}/4$ highest degree vertices $\tilde{A}'$ of $\tilde{A}$ in $\tilde{H}[\tilde{A},\tilde{B}']$. Then $\tilde{H}[\tilde{A}',\tilde{B}']$ must have at least $\tilde{c}/(4|\tilde{A}|)$ of the edges of $\tilde{H}[\tilde{A},\tilde{B}']$. So  $\tilde{H}[\tilde{A}',\tilde{B}']$ is feasible and since $n \geq |\tilde{A}|$
\begin{eqnarray}
|E(\tilde{H}[\tilde{A}',\tilde{B}'])| &\geq& \frac{\tilde{c}}{4|\tilde{A}|}|E(\tilde{H}[\tilde{A},\tilde{B}'])| \nonumber \\
&\geq& \frac{1}{4}|\tilde{A}|^{\frac{1-\alpha}{1+\alpha}-1}|E(\tilde{H}[\tilde{A},\tilde{B}'])| \nonumber \\ 
&\geq& \frac{1}{16}n^{\frac{-2\alpha}{1+\alpha}}MAX[\tilde{H},\tilde{c}].
\end{eqnarray}
Case 2: $|\tilde{A}| > \tilde{c}^{(1+\alpha)/(1-\alpha)}$: This case is the most challenging and we provide a graph transformation method to deal with it. Construct a graph $\tilde{H}^*$ from $\tilde{H}$ by replacing $\tilde{B}$ by $d$ copies $\tilde{B}_1,..,\tilde{B}_d$ of the vertices of $\tilde{B}$ with edges between $\tilde{A}$ and each $\tilde{B}_i$ copying the edges between $\tilde{A}$ and $\tilde{B}$. Thus if $\tilde{A}=\{a_i,i=1,..,r\}$, $\tilde{B}=\{b_i,i=1,..,s\}$ and $\tilde{B}_j=\{b_{ij},i=1,..,s, j=1,..,d\}$ then 
\begin{equation}
(a_i,b_j) \textrm{ is an edge of } \tilde{H} \textrm{ iff } (a_i,b_{jk}), k=1,..,d \textrm{ is an edge of } \tilde{H}^*.
\end{equation}
In $\tilde{H}^*$ set cost$(b_{ij})$=$(1/d)$cost$(b_{i})$ and so each cost$(b_{ij})$ is between 1 and 2 (see figure 1).
\begin{figure}
\begin{center}
\includegraphics[width=100mm]{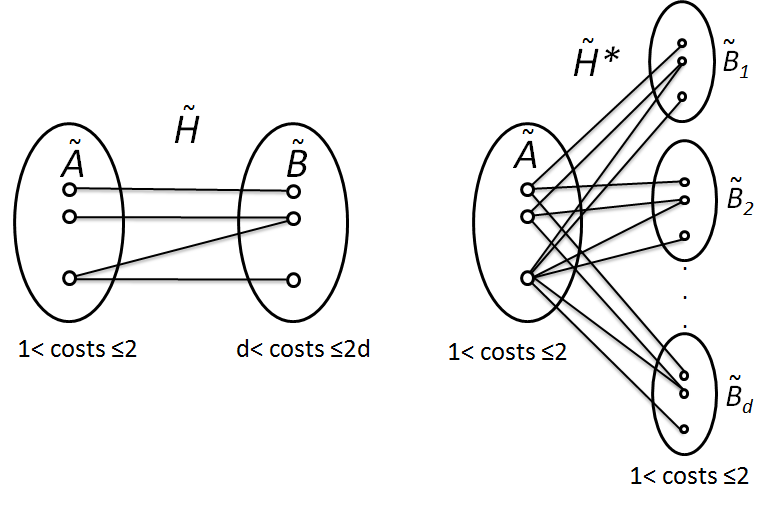}
\caption{\label{figure1.png} bipartite graphs $\tilde{H}$ and $\tilde{H}^*$}
\end{center}
\end{figure}
Now $MAX[\tilde{H}^*,\tilde{c}] \geq dMAX[\tilde{H},\tilde{c}]$ since any feasible subgraph of $\tilde{H}$ can be replicated $d$ times to a feasible solution of $\tilde{H}^*$. Use algorithm $M$ to approximate the DkS problem for $\tilde{H}^*$ with $k=\tilde{c}$ to within an approximation ratio of say $s$. Let $\tilde{H}^{**}[\tilde{A}',\tilde{B}'_1,\tilde{B}'_2,..,\tilde{B}'_d]$ be an induced subgraph corresponding to this approximation. Define $\delta_{ij}$ be the degree of $b_{ij}$ in $\tilde{H}^{**}$ (if $b_{ij}$ is not in $\tilde{H}^{**}$ then $\delta_{ij}=0$), and $\delta_i^*=max_j\{\delta_{ij}\}$. Now choose a subgraph of $\tilde{H}$ as follows. Select $B'$ to be those $b_i$ for which $\delta_i^*$ is among the $\tilde{c}/(4d)$ highest values among $\{\delta_j^*\}$. Then since $\cup \tilde{B}'_i$ contains at most $\tilde{c}$ vertices
\begin{equation}
|E(\tilde{H}[\tilde{A}',\tilde{B}'])| \geq \frac{1}{4d}|E(\tilde{H}^{**}[\tilde{A}',\tilde{B}'_1,\tilde{B}'_2,..,\tilde{B}'_d])| \geq \frac{1}{4ds}MAX[\tilde{H}^*,\tilde{c}].
\end{equation}
Now select $\tilde{A}''$ to be the $\tilde{c}/4$ highest degrees of $\tilde{A}'$ in $\tilde{H}[\tilde{A}',\tilde{B}']$. Since $|\tilde{A}'|$ is at most $\tilde{c}$, $|E(H[A'',B'])| \geq (1/4)|E(\tilde{H}[\tilde{A}',\tilde{B}'])|$ and so
\begin{equation}
|E(\tilde{H}[\tilde{A}'',\tilde{B}'])| \geq \frac{1}{16ds}MAX[\tilde{H}^*,\tilde{c}] \geq \frac{1}{16s}MAX[\tilde{H},\tilde{c}].
\end{equation}
$\tilde{H}[\tilde{A}'',\tilde{B}']$ is feasible since the total cost of vertices in each part is at most $\tilde{c}/2$. By the inequality defining this case 
\begin{equation}
\tilde{c} < |\tilde{A}|^{\frac{1-\alpha}{1+\alpha}} \leq n^{\frac{1-\alpha}{1+\alpha}}
\end{equation}
so $\tilde{H}^*$ has at most $nd$ vertices where
\begin{equation}
nd \leq \frac{1}{4}n\tilde{c} <  \frac{1}{4}n^{1+\frac{1-\alpha}{1+\alpha}} = \frac{1}{4}n^{\frac{2}{1+\alpha}}.
\end{equation}
Thus
\begin{equation}
s=O([\frac{1}{4}n^{\frac{2}{1+\alpha}}]^{\alpha+\epsilon})=O(n^{\frac{2\alpha}{1+\alpha}+\frac{2\epsilon}{1+\alpha}}).
\end{equation}
Similarly the time taken for $M$ to run is
\begin{equation}
O([\frac{1}{4}n^{\frac{2}{1+\alpha}}]^{1/\epsilon})=O(n^{\frac{2}{\epsilon(1+\alpha)}}).
\end{equation}
Noting that $2\epsilon/(1+\alpha)<2\epsilon$ and $2/(\epsilon(1+\alpha))<2/\epsilon$ the selection of $\tilde{H}[\tilde{A}'',\tilde{B}']$ through the use of $M$ provides an approximation ratio of at most $O(n^{2\alpha/(1+\alpha)+2\epsilon})$ to $MAX[\tilde{H},\tilde{c}]$ by inequality 6, and this is obtained within time $O(n^{2/\epsilon})$. This completes this case and also this class.

To complete the proof of the theorem we note that $G'^{MAX}[c]$ must share a vertex/edge profit of at least $1/[2(log_2n+1)^3+1]$ of $MAX[G',c]$ with at least one of the subgraphs $H$ in classes 1-5. Thus the largest profit found among the approximations to $MAX[H,c]$ over the subgraphs must approximate $MAX[G',c]$, and so $MAX[G,c]$, to within 
\begin{equation}
O((log_2n)^3n^{\frac{2\alpha}{1+\alpha}+2\epsilon})=O(n^{\frac{2\alpha}{1+\alpha}+2\epsilon+\frac{3log(log_2n)}{logn}}).
\end{equation}
For $n$ sufficiently large the term $3log(log_2n)/logn$ is less than $\epsilon$ and the desired approximation bound follows. The order of the run time can similarly be shown to be at most $O(n^{3/\epsilon})$.








\end{document}